\documentclass[pra,aps,longbibliography,showpacs,superscriptaddress,twocolumn,toc=flat,nofootinbib,groupedaddress]{revtex4-1}
\usepackage{graphicx}
\usepackage{amssymb}
\usepackage{amsmath}
\allowdisplaybreaks
\usepackage{physics}
\usepackage{amsfonts}
\usepackage{bm}
\usepackage{bbold}
\usepackage{verbatim}
\usepackage{ulem}
\usepackage[unicode=true,
 bookmarks=false,
 breaklinks=false,pdfborder={0 0 1},backref=false,colorlinks=true]
 {hyperref}
\hypersetup{
 linkcolor=[rgb]{0,0,1},citecolor=[rgb]{0,0,1},urlcolor=[rgb]{0,0,1}}
\usepackage{multirow}
\usepackage{color}
\usepackage{comment}
\usepackage{xcolor}
\usepackage{tabularx}

\allowdisplaybreaks
\newcommand{\circup}{%
\mathrel{\ooalign{$\circ$\cr\raise0.2ex\hbox{$\uparrow$}}}
}
\newcommand{\circdown}{%
\mathrel{\ooalign{$\circ$\cr\raise-0.1ex\hbox{$\downarrow$}}}
}

\begin{document}

\title{Realizing multi-orbital Emery models with ultracold atoms}

\author{Conall~McCabe}
\email{conall.mccabe@jila.colorado.edu}
\affiliation{JILA, National Institute of Standards and Technology, Center for Theory of Quantum Matter, and Department of Physics, University of Colorado, Boulder, CO, 80309, USA}
%\affiliation{Center for Theory of Quantum Matter, University of Colorado, Boulder, CO, 80309, USA}

\author{Jamie~Boyd}
\affiliation{JILA, National Institute of Standards and Technology, Center for Theory of Quantum Matter, and Department of Physics, University of Colorado, Boulder, CO, 80309, USA}

\author{Kaizhao~Wang}
\affiliation{JILA, National Institute of Standards and Technology, Center for Theory of Quantum Matter, and Department of Physics, University of Colorado, Boulder, CO, 80309, USA}

\author{Martin~Lebrat}
\affiliation{JILA, National Institute of Standards and Technology, Center for Theory of Quantum Matter, and Department of Physics, University of Colorado, Boulder, CO, 80309, USA}

%\author{Shiwei Zhang}
%\author{Andrew Millis}
%\author{authors}
%\affiliation{Flatiron}

\author{Cindy~Regal}
\affiliation{JILA, National Institute of Standards and Technology, Center for Theory of Quantum Matter, and Department of Physics, University of Colorado, Boulder, CO, 80309, USA}

\author{Adam~Kaufman}
\affiliation{JILA, National Institute of Standards and Technology, Center for Theory of Quantum Matter, and Department of Physics, University of Colorado, Boulder, CO, 80309, USA}

\author{Ana~Maria~Rey}
\affiliation{JILA, National Institute of Standards and Technology, Center for Theory of Quantum Matter, and Department of Physics, University of Colorado, Boulder, CO, 80309, USA}
%\affiliation{Center for Theory of Quantum Matter, University of Colorado, Boulder, CO, 80309, USA}

\author{Lukas~Homeier}
\email{lukas.homeier@jila.colorado.edu}
\affiliation{JILA, National Institute of Standards and Technology, Center for Theory of Quantum Matter, and Department of Physics, University of Colorado, Boulder, CO, 80309, USA}
%\affiliation{Center for Theory of Quantum Matter, University of Colorado, Boulder, CO, 80309, USA}

\date{\today}

\begin{abstract}
Strongly-correlated electrons in transition-metal oxides give rise to intriguing emergent phenomena, including high-temperature superconductivity in cuprates. While simplified one-band Hubbard models capture some aspects, explicitly describing the interplay of copper and oxygen orbitals -- as in the three-band Emery model -- is essential to capture the full phenomenology of cuprates. Quantum simulators based on ultracold atoms offer a promising route to study such systems in a controlled setting, but realizing realistic multi-orbital Hubbard models remains challenging. Here we propose an optical superlattice architecture that implements the three-band Emery model with ultracold fermions. By combining lattice beams with controllable interference, we engineer orbital degrees of freedom that reproduce key features of the cuprate band structure, while enabling independent control of orbital-dependent interactions and charge-transfer energy. We show that single-particle quantum walks can benchmark the resulting tight-binding model. Using determinant quantum Monte Carlo, we further investigate thermodynamic properties in the undoped regime and find a finite-temperature metal–insulator crossover accompanied by the onset of antiferromagnetic correlations accessible in current experiments. Finally, we apply a Hamiltonian learning protocol enabling to infer effective single-band Hubbard models from experimental realizations of Emery models. Our results provide a practical pathway to simulate multi-orbital Hubbard physics with quantum gas microscopes.
\end{abstract}

\maketitle

%%%%%%%%%%%%%%%%%%
%% Introduction
%%%%%%%%%%%%%%%%%%
\section{Introduction}
Many complex phenomena in condensed-matter physics emerge from interacting fermions. Quantum simulation in optical lattices provides a platform for their microscopic study~\cite{Gross2017} and has enabled impressive progress in realizing simplified models of strongly-correlated fermions~\cite{Kendrick2025,Bourgund2025,Chalopin2026}.
Under increasingly low temperature conditions~\cite{Kendrick2025,Chalopin2026,Xu2025}, controlled single-site microscopy~\cite{Gross2021} has enabled outstanding experimental progress on investigations of the single-band Hubbard model~\cite{Bohrdt2021}, which has been instrumental to our modern understanding of strongly-correlated electrons~\cite{Dagotto1994,Duan2003}. 
While these models reproduce many qualitative features of the phase diagram of high-$T_c$ superconductors, recent large-scale simulations~\cite{Qin2020}, however, indicate that such effective, simplified models need to be extended by additional terms, such as next-nearest-neighbor tunnelings~$t'$~\cite{Hirayama2018,Jiang2023,Xu2024}, to capture the low-energy physics. More fundamentally, this raises the question of the approximations required to capture the full range of correlated phases observed in cuprate superconductors~\cite{Keimer2015}, including the minimal set of electronic orbitals in the model Hamiltonian.

\begin{figure*}[t!!]
\centering
\includegraphics[width=\linewidth]{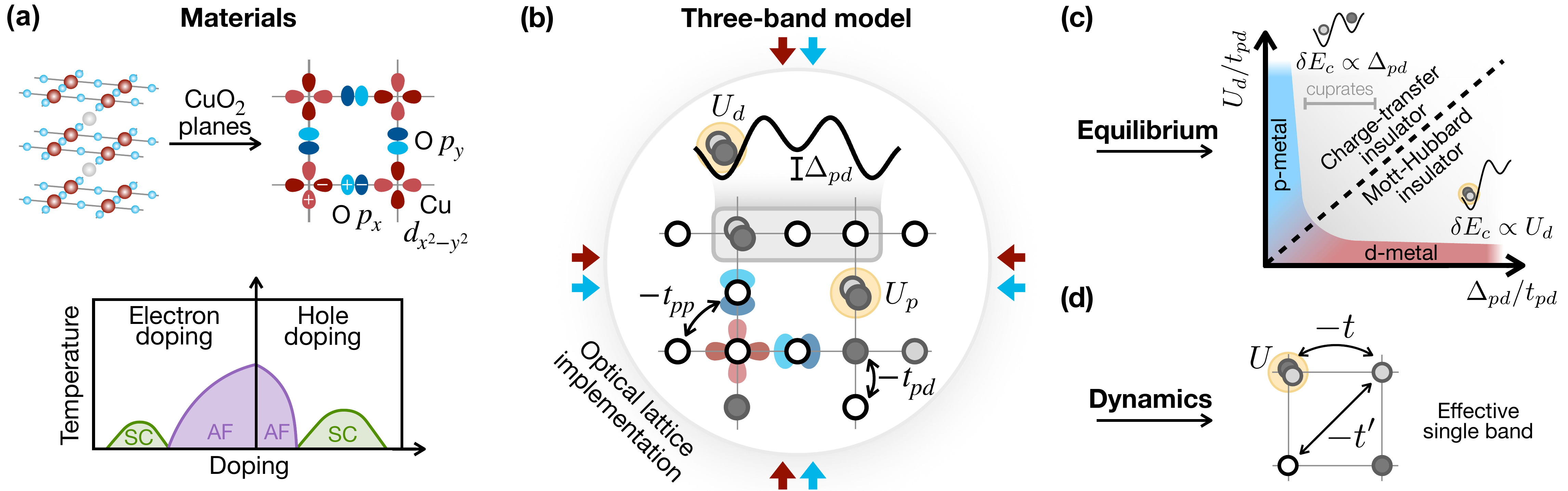}
\caption{\textbf{Emery model and its quantum simulation.} \textbf{(a)} In transition-metal oxides, such as cuprate superconductors, electronic correlations are dominated by three orbitals near the Fermi energy and are effectively confined to two-dimensional planes. The copper Cu (oxygen O) orbitals are located on the sites (links) of a square lattice, forming a Lieb lattice geometry. Their paradigmatic low-temperature phase diagram shows rich interplay between antiferromagnetic Mott insulators (AF) and mobile charges, including a high-T$_c$ superconducting (SC) dome whose differing size on either side highlights the particle-hole asymmetry. 
\textbf{(b)} We propose a bichromatic optical superlattice architecture with interfering and non-interfering lattice beams (sketched as blue and red arrows) to realize the three-band Emery model with ultracold fermions. Our scheme enables independent control over local Hubbard interactions~$U_{d,p}$ between two atoms in different spin states (shown as light and dark gray color) and charge-transfer energy~$\Delta_{pd}$. The dominant tunneling scale is set by $t_{pd}$ hybridizing neighboring $d$- and $p$-orbitals; our tight-binding model also features longer-range tunnel couplings, such as $t_{pp}$ present in cuprates. \textbf{(c)} The competition between orbital degrees of freedom is tuned by~$\Delta_{pd}/t_{pd}$, opening the possibility to explore metal-insulator transitions which we show are accessible with current experimental temperatures. The charge gap~$\delta E_c$ is determined by the hierarchy of $U_d$ and $\Delta_{pd}$, allowing one to study the different insulators known in transition-metal oxides.
\textbf{(d)} Time-evolution quench protocols can be used to learn effective single-band Hubbard model from experimental data and classical postprocessing.  }
\label{figure-1}
\end{figure*}

A canonical multi-orbital model for transition-metal oxides is provided by {\it the three-band Emery model}~\cite{Emery1987}. This model includes transition-metal $d$~orbitals --- those of Cu or Ni ions ---  with strong on-site repulsive interactions, $U_d$, which are coupled to oxygen $p$~orbitals with weaker interactions, $U_p$, see Fig.~\ref{figure-1}(a,b). 
The hybridization~$t_{pd}$ and orbital energy offset~$\Delta_{pd}$ introduce an additional tuning parameter absent in single-band models~\cite{Scalettar1991,Medici2009,Ponsioen2023}, where the competition solely arises from an interplay between tunneling~$t$ and on-site repulsion~$U$.
This extra degree of freedom naturally accounts for the particle–hole asymmetry observed in the cuprate phase diagram~\cite{Medici2009}, see Fig.~\ref{figure-1}(a) bottom, and enriches the competition between energy scales, as it can drive a metal–insulator transition~\cite{Zaanen1985,Imada1998}, see Fig.~\ref{figure-1}(c). However, the precise location of the phase boundary remains difficult to determine in numerical simulation~\cite{Kent2008,Karolak2010,Wang2012}. At the same time, $\Delta_{pd}$ sets the orbital occupancy and appears to anticorrelate with the superconducting critical temperature~$T_c$~\cite{Weber2012,Rybicki2016,Kowalski2021,Wang2023_Tc}, emphasizing that oxygen charges play an active role in the emergence of correlated phases.

Despite the impressive level of control  and advances in optical lattice technology~\cite{Tarruell2012,Greif2013}, which have enabled the realization of lattice geometries beyond single-band square-lattice Hubbard models, including  triangular~\cite{Xu2023,Mongkolkiattichai2023,Prichard2024,Lebrat2024} and Lieb lattices~\cite{Lebrat2024_Lieb,Taie2015} or mixed-dimensional models~\cite{Hirthe2023,Bourgund2025}, the realization of multi-orbital models such as the Emery model has remained  experimentally challenging. This is largely due to the difficulty of locally tuning Hubbard interactions and designing orbital-dependent band structures.

Here, we propose a way  to realize Emery models with ultracold fermions in optical superlattices. Our scheme relies on bichromatic lattices featuring interfering and non-interfering components. By tuning the intensities and interference phase, we describe how to engineer a tight-binding band structure with parameters close to those of cuprate superconductors, including independently tunable $U_d/U_p$ and charge-transfer gap~$\Delta_{pd}$.
We first benchmark the system in the simple case of non-interacting particles. We study quantum walks relevant to characterize the tight-binding parameters in optical lattices and discuss the absence of Fermi surface nesting due to band mixing. In the presence of interactions, this property leads to insulating states in the undoped system that are distinct from those of single-band Hubbard models.
To demonstrate the experimental feasibility for exploring the strongly-correlated regime, we perform determinant quantum Monte Carlo (DQMC) simulations to estimate the charge-ordering temperature, which  we show lies within the range accessible to current quantum simulators.
Finally, we propose a Hamiltonian learning protocol to obtain refined effective single-band Hubbard parameters from near ground-state quenches in the three-band model, illustrated in Fig.~\ref{figure-1}(d).

\section{Emery Model}
In cuprate superconductors, the electronic correlations of the two-dimensional copper-oxide planes are effectively described by a multi-orbital three-band Hubbard model (or Emery model) with Cu $d_{x^2-y^2}$ orbitals on the sites and ligand O $p_{x,y}$ orbitals on the links of a square lattice~\cite{Emery1987}, forming the Lieb lattice in Fig.~\ref{figure-1}(a).
This model is conveniently described in a hole representation.
In this representation, holes correspond to the \textit{absence} of electrons relative to a filled orbital configuration, such that the undoped, parent compound contains one hole per unit cell, primarily residing in the Cu $d_{x^2-y^2}$ orbital. The fermionic holes are annihilated (created) by $\hat{c}^{(\dagger)}_{j\sigma}$ on site~$j$ with spin~$\sigma$. In our scheme, the holes are emulated by fermionic atoms in an optical lattice.
The Hamiltonian in this hole picture, and under the tight-binding approximation, is given by 
\begin{align} \label{eq:Hamiltonian}
\begin{split}
\hat H =&
-\sum_{i<j}\sum_{\sigma} t_{ij} \left( \hat{c}^\dagger_{i\sigma} \hat{c}_{j\sigma} +\mathrm{h.c.}\right)
\\
&
+\sum_{\alpha\in\{p,d\}}\sum_{j_\alpha} U_\alpha \hat n_{j_\alpha\uparrow}\hat n_{j_\alpha\downarrow} +
\Delta_{pd}\sum_{j_p}\hat n_{j_p}.
\end{split}
\end{align}
Here, Cu (O) sites are labeled by~$j_d$ ($j_p$) and~$\hat{n}_{j_\alpha\sigma}=\hat{c}^\dagger_{j_\alpha\sigma}\hat{c}_{j_\alpha\sigma}$ is their occupation. We consider tunnelings between Cu-O sites~$t_{ij}=t_{pd},t'_{pd}, t''_{pd}$, between O-O sites~$t_{ij}=t_{pp},t'_{pp},t''_{pp}$ and between Cu-Cu sites~$t_{ij}=t_{dd}$, up to nearest-neighbor unit cells labeled by increasing distance.
In our convention~$t_{pd},t'_{pd},t_{pp},t''_{pp}>0$ and $t''_{pd},t'_{pp},t_{dd}<0$, see Appendix~\ref{app:gauge-trafo}. This structure of signs is consistent between cuprate compounds~\cite{Kent2008,Jiang2023} and our optical lattice implementation. The Hubbard interaction~$U_\alpha$ differs between Cu ($U_d$) and O sites ($U_p$). We choose the zero of energy such that the charge-transfer gap~$\Delta_{pd}$ is given by an additional offset on the O sites, see Fig.~\ref{figure-1}(b).
The specific tight-binding parameters in cuprate superconductors are currently not known precisely and vary across compounds, but generally take values of (in units of $t_{pd} \sim 1.2\,\mathrm{eV}$): $U_{d}/t_{pd} \sim 7$, $U_{p}/t_{pd} \sim 3\ldots5$, $\Delta_{pd}/t_{pd} \sim 2 \ldots 4$, $t_{pp}/t_{pd} \sim 0.3$, $-t'_{pp}/t_{pd} \sim 0.15$, $-t_{dd}/t_{pd} \sim 0.1$, see e.g. Refs.~\cite{Emery1987,Kent2008,Weber2012,Vitali2019,Jiang2023}, with decreasing tunneling amplitudes for larger distances.

\begin{figure}[t!!]
\centering
\includegraphics[width=\linewidth]{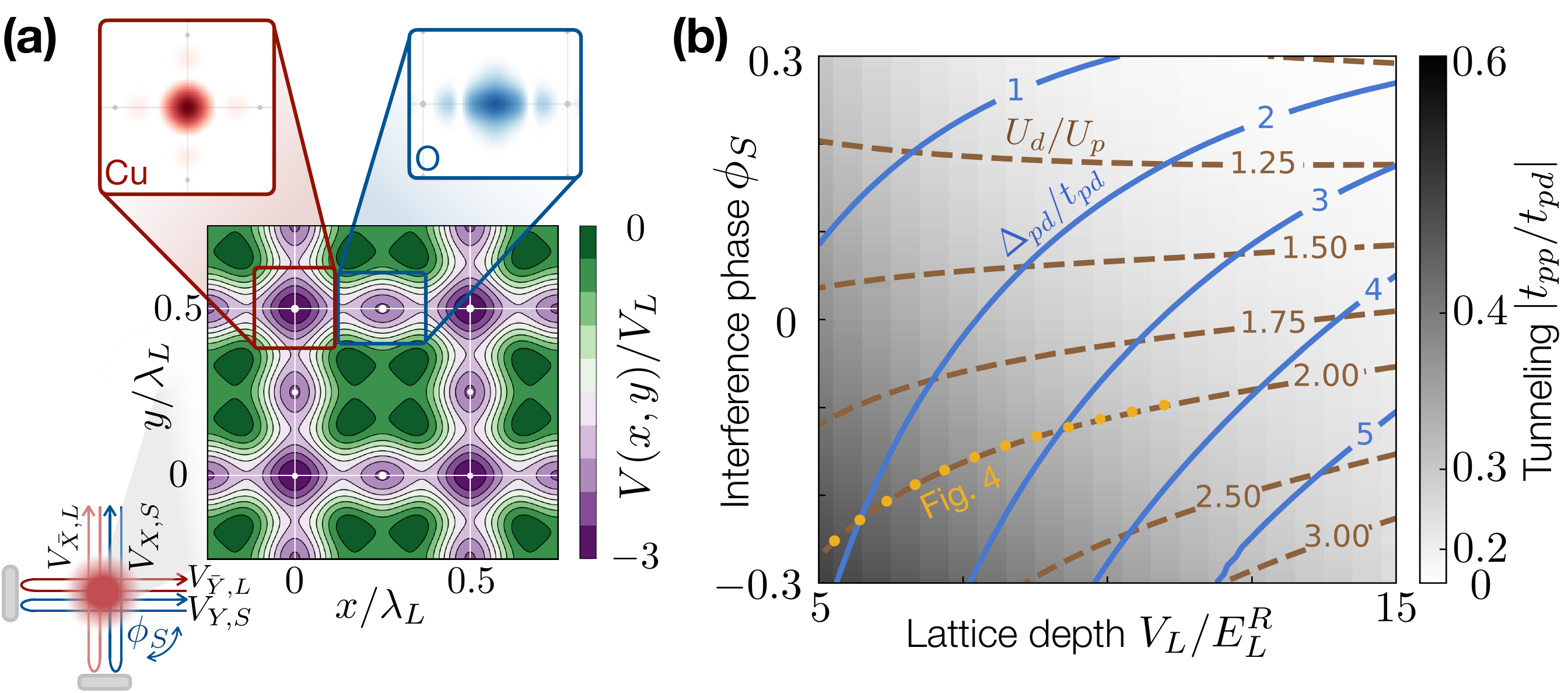}
\caption{\textbf{Optical lattice implementation.} \textbf{(a)} Bichromatic scheme with interfering short lattice (blue) and non-interfering long lattice (red). Center: Optical lattice potential for $\phi_S=0.1$ and~$V_S/V_L=1.25$. The potential minima (purple) form the underlying Lieb lattice. Inset: Probability density of maximally-localized Wannier orbitals on the Cu site (red) and O sites (blue) plotted on a log-scale. \textbf{(b)} Accessible tight-binding parameters varying overall lattice depth and keeping~$V_S/V_L=1.25$ fixed with contour lines for $U_d/U_p$ (dashed brown lines) and $\Delta_{pd}/t_{pd}$ (solid blue lines). The change in band structure, such as~$|t_{pp}/t_{pd}|$, is indicated by the gray background. The absolute tunneling amplitude along the dotted orange contour is $t_{pd}/E^{R}_{L} = 0.62 \ldots 0.68$.}
\label{figure-2}
\end{figure}

In materials, the $d$-wave and $p$-wave symmetry of orbitals determines the tunneling sign structure leading to a staggering of tunnelings between sites belonging to unit cells of opposite sublattices. Since the Lieb lattice is bipartite, we can apply a gauge transformation~$\hat{c}_{j\sigma} \rightarrow -\hat{c}_{j\sigma}$ for all~$j$ in one sublattice. The transformation equalizes the sign of the relevant tunneling amplitudes we consider, see Fig.~\ref{figure-1}(b) and Appendix~\ref{app:gauge-trafo}; while this transformation is trivial, it is necessary in order to map the $p$-$d$~model to tight-binding descriptions in optical lattice potentials. The choice of gauge has to be acknowledged in observables that transform non-trivially.

%%%%%%%%%%%%%%%%%%
%% Optical Lattice Implementation
%%%%%%%%%%%%%%%%%%
\section{Optical Lattice Implementation}
\label{sec:Implementation}
We propose a 2D optical superlattice architecture composed of (i) a short-wavelength~$\lambda_S$, interfering 2D lattice, originating from two orthogonal, retro-reflected beams with same frequency, actively stabilized relative phase~$\phi_S$ and equal 1D lattice potentials~$V_{X,S}=V_{Y,S}=V_{S}$, and (ii) a long-wavelength~$\lambda_L=2\lambda_S$ retro-reflected non-interfering lattice with potentials~$V_{\bar{X},L}=V_{\bar{Y},L}=V_{L}$ originating from frequency detuned laser beams~\cite{Tarruell2012,Xu2023}, see Fig.~\ref{figure-2}(a) bottom. A tight potential in the perpendicular direction (not shown here) confines the atoms in a 2D plane. The resulting attractive optical potential on the atoms in the $x$-$y$ plane is given by
\begin{align}
\begin{split}
    V(x,y) = -\sum_{A=S,L}&V_A\big[  \cos^2 (k_A x) + \cos^2 (k_A y)\big] \\
    + 2&V_S\sin \phi_S \cos (k_S x) \cos (k_S y) - c.
\end{split}
\end{align}
with~$k_{A}=2\pi/\lambda_{A}$ and we define the recoil energy\footnote{For $\lambda_L=2\lambda_S =2060\,\mathrm{nm}$, our superlattice wavelengths chosen to reduce off-resonant scattering for fermionic~${}^6\mathrm{Li}$, the recoil energies are~$E^R_S=h\cdot31.1\,\mathrm{kHz}$ and~$E^R_L=h\cdot7.8\,\mathrm{kHz}$.} for atoms with mass~$m$ as~$E^R_A=h^2/2m\lambda_A^2$. We set the constant offset~$c=\mathrm{max}V(x,y)$. For $V_L/V_S \simeq 0.5\ldots1.5$ and $\phi_S/\pi \simeq -0.1\ldots0.1$, the potential generates a Lieb lattice landscape~\cite{DiLiberto2016,Flannigan2021} wherein each local minima represents a site illustrated in purple in Fig.~\ref{figure-2}(a).  

To gain intuition, we consider the case of maximal interference\footnote{We note that the particular case of maximal interference~$\phi_S = \pi/2$, $V_S/V_L \sim 1/4$ was considered in Refs.~\cite{DiLiberto2016,Flannigan2021}. }, when $\phi_S= \pi / 2 $. $V_L$ generates a square lattice with unit length $a=\lambda_L/2$, forming the Cu \textit{d} sites, while $V_S$ generates a square lattice rotated by $45^{\circ}$ with spacing $\lambda_L/(2\sqrt{2})$, forming the O \textit{p} sites. The difference in lattice potentials~$V_S - V_L$ leads to an offset between $d$- and $p$-sites, i.e., a charge-transfer gap~$\Delta_{pd}$. 
The model parameters, in particular the relative Hubbard interaction~$U_d/U_p$ and charge-transfer gap~$\Delta_{pd}$, are tuned by varying the phase~$\phi_S$, the relative lattice depth~$V_S/V_L$ and the overall lattice depth~$V_L$, see Fig.~\ref{figure-2}(b). By following the contour lines, those parameters can be tuned independently through regimes relevant to cuprates~$U_d > U_p$ and~$\Delta_{pd}/t_{pd}=2...5$, while keeping the band structure approximately constant, i.e., a weakly varying ratio~$t_{pp}/t_{pd}$ and absolute scale of tunneling amplitudes, see Appendix~\ref{app:Wannier}. The absolute interaction scale~$U_d$ is tuned by an atomic scattering resonance. 

To precisely obtain the tight-binding parameters, we compute the Wannier functions of this non-separable lattice potential via a steepest-descent algorithm to maximally localize orbitals by mixing the three lowest bands~\cite{Walters2013,Marzari1997}, see Fig.~\ref{figure-2}(a) top and Appendix~\ref{app:Wannier}. All tunneling and interaction parameters of the tight-binding Hamiltonian~\eqref{eq:Hamiltonian} are obtained \textit{ab initio} from overlap integrals of two and four Wannier functions, respectively. While in conventional single-band optical lattice systems, the separability of the lattice potential prohibits diagonal tunneling, the band mixing applied here breaks the separability of the Wannier functions for any~$\phi_S$, yielding non-zero~$t_{pp}$.

\begin{figure*}
    \centering
    \includegraphics[width=\linewidth]{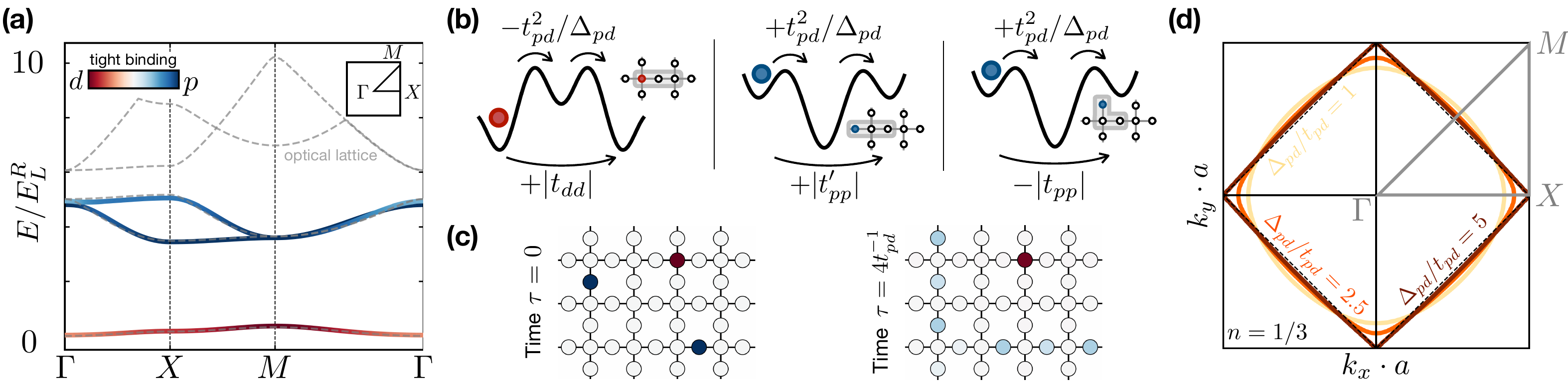}
    \caption{\textbf{Band structure.} \textbf{(a)} Example band structure of the optical lattice potential for $V_L = 15E^R_L$, $V_S/V_L=1.25$ and $\phi_S = 0$ showcasing three low-lying bands with $\Delta_{pd}/t_{pd}= 4.2$. The dashed lines are the exact structure from the optical lattice potential and the bold lines represent our tight-binding model, where the color indicates the Cu and O character of each band. \textbf{(b)} Direct and virtual tunneling process of single particles initialized in Cu or O sites interfere, which can be used to precisely benchmark the optical lattice parameters. \textbf{(c)} We show the quantum walk of particles starting in three different orbitals. A particle starting on the Cu site (red) is only weakly dispersing due to destructive interference. A particle starting in an O site (blue) spreads anisotropically. \textbf{(d)} Fermi surface of non-interacting fermions at fillings corresponding to the undoped compound for $\Delta_{pd}/t_{pd}=1,2.5,5$ (light to dark). Hybridization between $d$- and $p$-orbitals deforms the Fermi surface and suppresses nesting. }
    \label{figure-3}
\end{figure*}

To validate our tight-binding description and to gain confidence to have found the maximally-localized set of Wannier functions, we compare the resulting band structure, extracted from computing Wannier function overlaps, to the bands obtained from solving the Schr\"{o}dinger equation of a single-particle in the optical potential. In Fig.~\ref{figure-3}(a) we demonstrate excellent agreement with the tight-binding model by including tunneling within ($t_{pd},t_{pp},t'_{pp}$) and between nearest-neighbor unit cells ($t_{dd},t''_{pp},t'_{pd},t''_{pd}$). 
%The relatively extended wings of Wannier functions on the O sites generate tunnelings~$|t'_{pp}| \lesssim 0.5 t_{pd}$ and $|t_{pp}| \approx |t_{dd}| \approx 0.1t_{pd}$. 

For interacting atoms, collisions induce transitions to non-Emery bands, i.e., the fourth and higher bands of the optical lattice. These collisions determine the heating rate and feasibility of our scheme. 
The transition rates to the non-Emery bands can be estimated perturbatively and depend on spatial Wannier function overlaps, band gaps and the chosen scattering length, thereby setting an upper bound on the absolute interaction scale. In Fig.~\ref{figure-3}(a), the two upper gray dashed lines correspond to the fourth and fifth band of the optical lattice. While the energy gaps between the fourth, and the second and third band are small compared to the bandwidths of Emery bands, their coupling matrix element is also small, because the fourth band is largely located around the ``fake site'' at $x=y=0.25\lambda_L$ in Fig.~\ref{figure-2}(a) (center of plaquette in Lieb lattice). 
%Therefore, the dominating coupling matrix element to these bands is obtained from the lowest $d$-band, for which the energy gaps are large, suppressing collisional transitions. 
We systematically estimate the collisional rates in Appendix~\ref{app:Wannier} and find them to be smaller than relevant energy scales in the system, showing the feasibility of our scheme for the lattice depth and corresponding parameters we consider throughout the manuscript.

In our lattice setup, typical nearest-neighbor tunneling scales~$t_{pd} \approx 0.15E^R_S = 0.6E^R_L$ (for $V_L=10E^R_L$, $V_S/V_L=1.25$ and~$\phi_S=-0.1$) are significantly larger than those obtained in comparable single-band models. However, the relevant tunneling scale is set by the effective tunneling~$t_{\rm eff} \sim t^2_{pd}/\Delta_{pd}$, which controls the onset of incompressible behaviour. For typical cuprates the effective tunneling is~$t_{\rm eff} \approx 0.3 t_{pd}$. Since current single-band Fermi-Hubbard quantum simulators routinely reach temperatures of~$k_B T/t \lesssim 0.2$~\cite{Xu2023,Lebrat2024,Prichard2024,Chalopin2026}, where~$t$ is the characteristic nearest-neighbor tunneling amplitude in a square lattice, we expect that a finite-$T$ crossover, reminiscent of the metal-insulator transition, is experimentally observable in the three-band model as discussed in Sec.~\ref{sec:DQMC}.

To reach even lower temperatures, the cooling protocols demonstrated in Ref.~\cite{Xu2025} can be naturally integrated into our bichromatic lattice architecture. 
We propose preparing a band insulator in a long-wavelength lattice with spacing $\lambda_L/\sqrt{2}$, generated by the interfering lattice. Upon adiabatic ramping into the proposed lattice, this state maps onto a filling of one fermion per unit cell, corresponding to an undoped system.
Recent studies show that band insulators can be prepared with high fidelity while largely preserving low entropy during the ramp. This approach is particularly favorable in our setup, where the large bare tunneling scale~$t_{pd}$ -- compared to single-band Hubbard models -- supports adiabaticity, as the relevant spin gap of the many-body system is set by superexchange processes~$J$, see Appendix~\ref{app:superexchange}.

%%%%%%%%%%%%%%%%%%
%% Non-interacting fermions
%%%%%%%%%%%%%%%%%%
\section{Non- and weakly-interacting fermions}
The single-atom addressability allows an isolated characterization of the engineered multi-orbital band structure via quantum walks~\cite{Weitenberg2011,Preiss2015,Young2022,Wei2023}. To this end, we study a single-particle dispersing under a Hamiltonian with $t_{pp} = -t'_{pp} = -t_{dd}=0.1t_{pd}$ and $\Delta_{pd}= 10t_{pd}$, starting with a localized atom prepared in different initial orbitals, as shown in Fig.~\ref{figure-3}(b,c). 
We find localization of the particle starting on the Cu site as well as a spatially anisotropic spreading for a particle starting on the O site, for the timescales we consider. We can explain this behaviour perturbatively by analyzing resonant tunneling processes, i.e., between Cu-Cu and O-O, see Fig.~\ref{figure-3}(b). A particle on the Cu site traverses to its neighboring Cu site via a direct tunneling term of amplitude~$+|t_{dd}|$ that interferes destructively with a virtual process~$-t^2_{pd}/\Delta_{pd}$, leading to perfect cancellation within our parameters. A particle on the O $p_x$ orbital couples to the nearest-neighbor~$p_x$ orbital by constructive interference of~$+|t'_{pp}|$ with the virtual tunneling process~$+t^2_{pd}/\Delta_{pd}$. However, the coupling to a neighboring~$p_y$ orbital experiences destructive interference between the second-order process with~$-|t_{pp}|$, constraining its motion to one dimension.
For parameters away from the perfect destructive interference, a particle can disperse in all directions~\cite{Wei2023}.

Next, we consider weakly-interacting fermions at finite densities with fillings corresponding to undoped cuprates. In the undoped compounds (corresponding to half filling in effective single-band Hubbard models) the system has one hole per unit cell, i.e., $\langle \hat{n}_{j_d} + 2 \hat{n}_{j_p}\rangle=1$ or filling~$n=1/3$. In the nearest-neighbor single-band square-lattice Hubbard model, the Fermi surface exhibits perfect nesting, leading to antiferromagnetic correlations and opening of a charge gap at arbitrarily small interactions~\cite{Imada1998,Fradkin2013}. Here, we investigate the nesting properties of the Fermi surface in the three-band Hubbard model by varying~$\Delta_{pd}$ in our optical lattice setup. For large~$\Delta_{pd} \gg t_{pd}$, the lowest band has negligible orbital mixing and is dominated by the Cu-Cu tunneling with dispersion~$ \propto t_{dd}\left(\cos(k_x a) + \cos(k_y a)\right)$. In this limit, the undoped system recovers the nested Fermi surface of the single-band Hubbard band, see Fig.~\ref{figure-3}(d). As the energy gap, set by the charge transfer gap and the bandwidths~$\propto t_{pp}, t_{dd}$, becomes comparable to~$t_{pd}$, band mixing deforms the Fermi surface and destroys perfect nesting. Nevertheless, the remaining near-nesting is still expected to enhance the tendency toward an antiferromagnetic insulating state at finite interactions~\cite{Fradkin2013}. 

\begin{figure}[t!!]
\centering
\includegraphics[width=\linewidth]{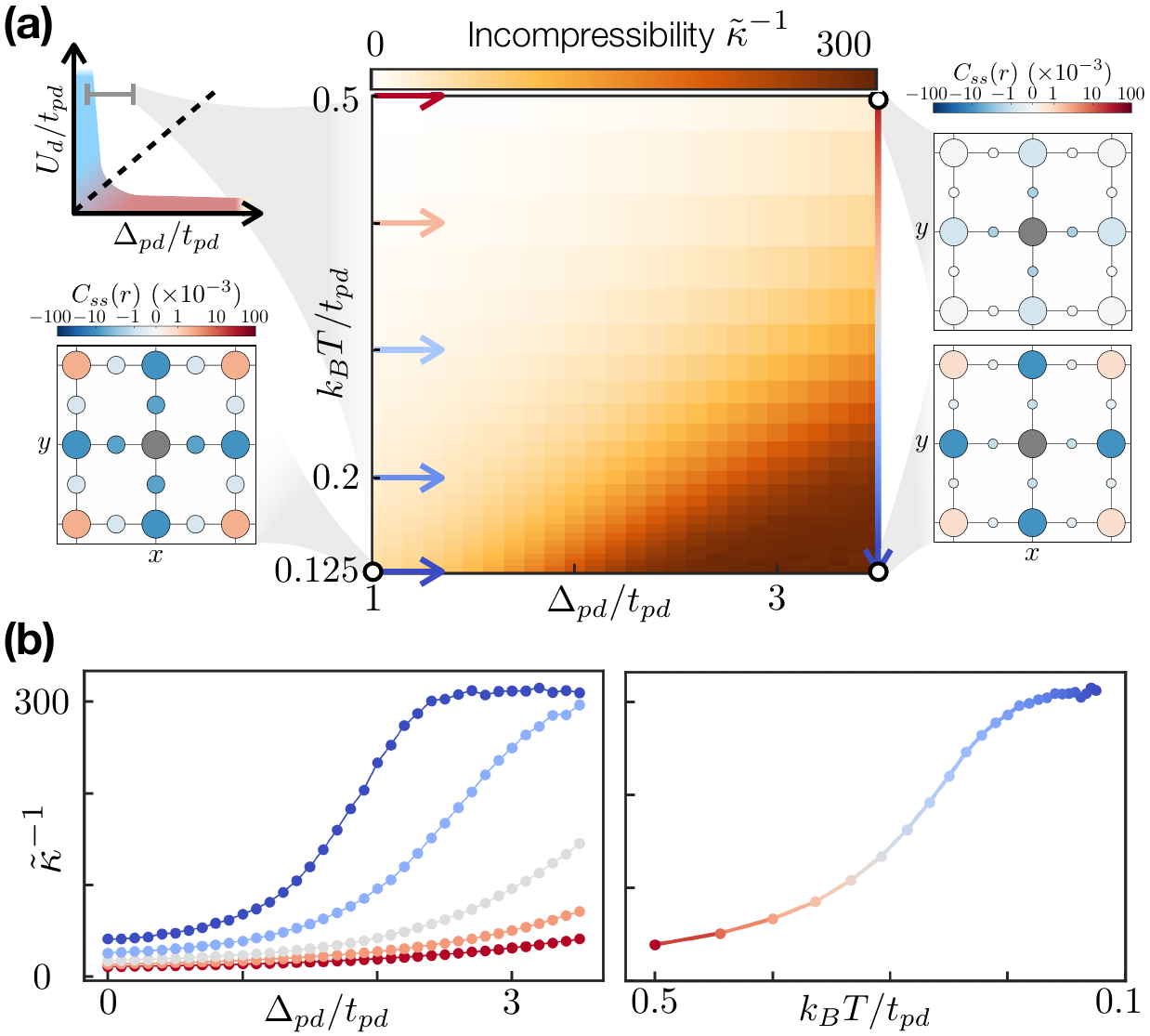}
\caption{\textbf{Finite-$T$ charge and spin correlations.} \textbf{(a)} Center: Dimensionless incompressibility $\tilde{\kappa}^{-1}$ obtained from DQMC on an $8 \times 8 $ system for $U_{d} = 7, U_{p} = 3.5, t_{pp} = 0.2, t_{pp}' = t_{dd} = 0.1$ (units of $t_{pd})$, approximately obtained from the parameter path in Fig.~\ref{figure-2}(b). For small~$\Delta_{pd}/t_{pd}$ the system remains compressible for all accessible temperatures. As we increase~$\Delta_{pd}/t_{pd}$, an incompressible phase appears reminiscent of the charge-transfer insulator in Fig.~\ref{figure-1}(c). Subpanels: We observe the onset of antiferromagnetic, exponentially decaying spin-spin correlations~$C_{ss}(r)$ at low temperatures plotted on a semilog scale (linear scale for $|C_{ss}(r)|\leq 10^{-3}$). The area of the sites visualizes the relative local occupation of $p$ and $d$ orbitals.
\textbf{(b)} Left: Cut of $\tilde{\kappa}^{-1}$ for $k_B T/t_{pd}=0.125,0.2,0.3,0.4,0.5$ (blue to red). Right: Cut for $\Delta_{pd}/t_{pd} = 3.5$. Current experiments are expected to access temperatures of~$k_B T \lesssim 0.2 t_{pd}$, allowing them to reach the correlated regime in multi-orbital systems. }
\label{figure-4}
\end{figure}
%%%%%%%%%%%%%%%%%%
%% Metal-to-Insulator Transition at finite-$T$
%%%%%%%%%%%%%%%%%%
\section{Metal-to-Insulator crossover at finite-$T$} \label{sec:DQMC}
At strong coupling, $U_d \gg t_{ij},U_p$, the formation of an insulator is enriched by an interplay between different orbitals, the orbital admixing is set by the competition between charge-transfer gap~$\Delta_{pd}$ and tunneling~$t_{pd}$. To characterize this, we consider the charge gap~$\delta E_{c}$ of the undoped compounds, following the Zaanen-Sawatzki-Allen classification of transition-metal oxides~\cite{Zaanen1985}. 

In the limit~$\Delta_{pd} > U_d$, holes (or fermionic atoms in our mapping) are predominantly located on the Cu sites, forming an antiferromagnetic Mott insulator with one hole (fermion) per Cu site and only small fluctuations to O sites. Due to the dominant charge-transfer energy, an additional hole occupies the Cu sites, which maps to a doubly occupied site in the experimental lattice scheme. Hence the charge gap~$\delta E_c \propto U_d$ is determined by the interactions, identifying this phase as Mott-Hubbard insulator, see Fig.~\ref{figure-1}(c) bottom right.

In the limit~$U_d > \Delta_{pd} > t_{pd}$, in the undoped compound, fermions still predominantly occupy the Cu sites despite larger hybridization with O sites. This undoped background also exhibits antiferromagnetic correlations originating from superexchange~$J$, however with substantial renormalization of~$J$ arising from the mobile $p$-bands~\cite{Eskes1993}. Strong interactions suppress doublons on the Cu site, such that an additional charge is transferred to the O site with the energy cost of~$\delta E_{c} \propto \Delta_{pd}$. This phase is identified as charge-transfer insulator, see Fig.~\ref{figure-1}(c) top left. At low energies, the additional charge forms a spin singlet with the adjacent Cu site, the so-called Zhang-Rice singlet~\cite{ZhangRice1988}. In hole-doped cuprates, this composite object is the mobile low-energy degree of freedom and corresponds to the holes in effective single-band Hubbard models~\cite{Tjeng1997}.

As the charge-transfer energy~$\Delta_{pd}$ is lowered even further and becomes comparable to the bandwidth of the itinerant oxygen band, a metallic state is realized, accompanied by a depletion of antiferromagnetic order~\cite{Vitali2019}. As the transition occurs from the interplay between mobile charges and strong interactions, the transition is a strongly-correlated phenomenon of the multi-orbital system. It is well-established that cuprates exist in the charge-transfer regime~\cite{Nuecker1988,Chen1991,Pellegrin1993}, but close to this metal-insulator transition with strongly hybridized $p$-$d$ bands~\cite{Imada1998,Wang2012,Vitali2019,Reaney2025}. 

However, a fully self-consistent determination of this transition in density-function theory and dynamical mean-field theory (DFT+DMFT) is challenging due to a double counting of on-site correlations and systematic errors in computing phase boundaries~\cite{Kent2008,Karolak2010,Wang2012}. This prohibits finding accurate estimates of~$\Delta^{\rm critical}_{pd}$, despite its importance highlighted by phenomenological~\cite{Rybicki2016,Wang2023_Tc} and numerical observations~\cite{Weber2012,Kowalski2021} which anticorrelate~$\Delta_{pd}$ with superconducting critical temperature~$T_c$. This anticorrelation suggests a picture where highest $T_c$'s are achieved when the system establishes sufficiently strong antiferromagnetic correlations while maintaining mobility of charges in the oxygen band~\cite{OMahony2022,Vitali2019}.

Here, we investigate the feasibility of observing the metal-insulator crossover in experimental realizations at realistic temperatures, where the relevant energy scales differ substantially from those of the single-band Hubbard model.
In the square-lattice single-band Hubbard model, the characteristic energy scale is the nearest-neighbor tunneling~$t$, with temperatures of $k_B T/ t \leq 0.2$ now routinely achieved~\cite{Xu2025,Kendrick2025,Chalopin2026}. In our system, however, charge dynamics are governed by an effective scale~$t_{\rm eff} \sim t^2_{pd}/\Delta_{pd} < t_{pd}$, while experimentally accessible temperatures are determined by the bare tunneling~$t_{pd}$, with ~$k_BT/t_{pd} \leq 0.2$\footnote{Using our lattice parameters, i.e., $t_{pd}=0.6E^R_L$, $\lambda_L=2060\,\mathrm{nm}$ and~$^{6}\mathrm{Li}$, the temperature in absolute units is~$k_BT=0.2  t_{pd} = 45\,\mathrm{nK}$.}. At these temperatures we expect a finite-$T$ crossover, reminiscent of the metal-insulator transition, to be experimentally observable, accompanied by the appearance of short-ranged and weak spin correlations.

We employ numerical DQMC simulations to identify the required temperatures, using the \texttt{SmoQyDQMC} toolkit~\cite{SmoQyDQMC.jl,SmoQyDQMC.jl_Codebase}.
We consider conventional cuprate parameters, which are accessible in our experimental implementation, $U_{d}/t_{pd} = 7$, $U_{p}/t_{pd} = 3.5$, $t_{pp}/t_{pd} = 0.2$, $t_{pp}'/t_{pd} = t_{dd}/t_{pd} = -0.1$, and we vary the charge-transfer gap~$\Delta_{pd}/t_{pd}=1 \ldots 3.5$, see dotted orange contour in Fig.~\ref{figure-2}(b).
To probe the onset of an insulating state, we compute the dimensionless compressibility $\tilde{\kappa} = \beta^{-1} (\partial n/\partial\mu)$, where~$\beta$ is the inverse temperature, $n$ the density. This quantity is directly accessible in quantum gas microscopes through local charge fluctuations~\cite{Duarte2015,Greif2016,Kendrick2025},
\begin{align} \label{eq:incompressibility}
    \tilde{\kappa} &= \frac{1}{\mathcal{N}}\left( \langle \hat{N} \rangle^2 - \langle \hat{N} \rangle^2 \right),
\end{align}
where $\hat{N} =\sum_{j,\sigma}\hat{n}_{j\sigma}$ and~$\mathcal{N}$ is the total number of lattice sites; in our simulations~$\mathcal{N}=192$ corresponding to a system size of~$8 \times 8$ unit cells with periodic boundary conditions.

In Fig.~\ref{figure-4}(a), we plot the incompressibility~$\tilde{\kappa}^{-1}$ as a function of the charge-transfer gap~$\Delta_{pd}$ for temperatures down to $k_B T/t_{pd}\geq 1/8$, where the fermionic sign remains well controlled. 
At large~$\Delta_{pd}/t_{pd}$ and low temperatures, we observe an incompressible regime, which we identify as a charge-transfer insulator. 
We define a crossover temperature~$T_{\rm cross}$ via a contour of constant~$\tilde{\kappa}^{-1}_{\rm const}$, which approximately follows the scaling~$\tilde{\kappa}^{-1}_{\rm const}\propto \Delta_{pd}/k_B T_{\rm cross}$. From thermal activation of the charge gap, we expect~$\tilde{\kappa}^{-1} \sim e^{\delta E_c / k_B T}$, implying that contours of constant~$\tilde{\kappa}^{-1}$ correspond to $k_B T_{\rm cross} \propto \delta E_c$. The observed scaling $k_B T_{\rm cross} \propto \Delta_{pd}$ therefore indicates that the charge gap scales as $\delta E_c \propto \Delta_{pd}$, consistent with charge-transfer insulating behavior.

As~$\Delta_{pd}/t_{pd}$ is reduced, the incompressibility significantly decreases even at low temperatures, consistent with a crossover toward a metallic state at finite temperature. This behavior reflects the closing of the charge gap as $\Delta_{pd}$ approaches the bandwidth scale, and the onset of itinerant oxygen charges. The strong variation of the compressibility provides a clear experimentally accessible signature for current ultracold-fermions experiments, see Fig.~\ref{figure-4}(b).

In Fig.~\ref{figure-4}(a) subpanels, we show the spin-spin correlation between a Cu site and its surrounding Cu and O sites~$C_{ss}(\mathbf{r})= L^{-2}\sum_{j_d}\langle \hat{S}^z_{j_d} \hat{S}^z_{j_d+\mathbf{r}} \rangle$, where~$L$ is the linear system size (unit cells),~$\hat{S}^z_{j}=(\hat{n}_{j\uparrow}-\hat{n}_{j\downarrow})/2$ is the $z$~projection of the spin operator acting at site $j$, and $\mathbf{r}=(x,y)/(a/2)$ is the displacement vector in units of the short lattice constant. In our plot, the area of each site visualizes the relative local occupations~$\langle \hat{n}_{p,d} \rangle$. In the charge-transfer insulating regime, Fig.~\ref{figure-4}(a) right, we find that antiferromagnetic correlations between Cu sites are developing as temperature is lowered.
Further, despite the potential offset~$\Delta_{pd}>0$, the strong hybridization~$t_{pd}$ leads to a finite occupation of $p$~orbitals and the formation of Zhang-Rice-like antiferromagnetic correlations between Cu and O sites.

As the charge-transfer energy is lowered, Fig.~\ref{figure-4}(a) left, the occupation shifts into the $p$-orbitals. In this regime, antiferromagnetic superexchange~\cite{Eskes1993} is dominated by~$J \approx 8t_{pd}^4/\Delta_{pd}^2(2\Delta_{pd}+U_p)$, originating from Pauli blocking, see Appendix~\ref{app:superexchange}. Therefore, the lowest temperature~$k_B T/t_{pd}=0.125$ is below the perturbative superexchange scale, enhancing the antiferromagnetic Cu-Cu correlations compared to larger~$\Delta_{pd}/t_{pd}$, as can be seen from the signal on the diagonals, though they remain exponentially decaying~\cite{Vitali2019}.

%%%%%%%%%%%%%%%%%%
%% Learning effective single-band models
%%%%%%%%%%%%%%%%%%
\section{Learning effective single-band models}
Recently, there has been renewed interest in constructing accurate effective single-band models for the low-energy physics of cuprate superconductors~\cite{Qin2020,Xu2024}. While early approaches relied on perturbative approaches~\cite{ZhangRice1988,Eskes1993}, recent work has instead used large-scale ground-state simulations to numerically downfold three-band models into effective single-band descriptions~\cite{Jiang2023,Lange2025}. This raises the question of whether downfolding can instead be achieved from quench experiments, accessible in quantum simulators, without input from large-scale numerical simulations.

Here, we adapt protocols for Hamiltonian learning based on quantum quenches~\cite{Bairey2019,Li2020,Ott2024} to infer effective single-band Hubbard models directly from experimentally accessible correlations. In these protocols, initial (product) states evolve under the system's Hamiltonian to time~$\tau$, after which correlators~$\langle \hat{\mathcal{O}}_j \rangle_\tau$ corresponding to an ansatz Hamiltonian~$\hat{H}' = \sum_{j} \alpha_j \hat{\mathcal{O}}_j$ are measured. 
As proposed in Ref.~\cite{Li2020}, the \textit{learned} Hamiltonian parameters~$\{ \alpha_j \}$ (up to a global factor) are obtained by solving a set of linear equations resulting from these measurements.

In our case, we choose an ansatz single-band $t-t'-U$~Hamiltonian
\begin{align}
\begin{split}
    \hat{H}' = &-t\sum_{\langle i_d,j_d \rangle,\sigma} \left(\hat{\tilde{c}}^\dagger_{i_d\sigma}  \hat{\tilde{c}}_{j_d\sigma} +\mathrm{h.c.} \right) \\ &-t'\sum_{\langle\langle i_d,j_d \rangle\rangle,\sigma} \left(\hat{\tilde{c}}^\dagger_{i_d\sigma}  \hat{\tilde{c}}_{j_d\sigma} +\mathrm{h.c.} \right) \\
    &+ U \sum_{j_d} \hat{\tilde{n}}_{j_d\uparrow}\hat{\tilde{n}}_{j_d\downarrow},
\end{split}
\end{align}
with dressed fermionic operators~$\hat{\tilde{c}}^\dagger_{j_d\sigma}$ defined on sites~$j_d$ of a single-band Hubbard model, that describe the $p$-$d$ hybridization of fermions in the three-band model~\cite{Jiang2023,Lange2025}. For sufficiently large~$\Delta_{pd}/t_{pd}$ and undoped systems, where fermions predominantly occupy Cu sites, we assume $\hat{c}^\dagger_{j_d\sigma} \approx \hat{\tilde{c}}^\dagger_{j_d\sigma}$.

Our protocol aims to learn \textit{effective} parameters~$\{ \alpha_j \} = \{t,t',U\}$, that capture the low-energy physics of the Emery model~\eqref{eq:Hamiltonian}. To do so, we quench an initial state in the three-band system and measure the corresponding correlators of the single-band model, e.g., we measure $\langle \hat{\tilde{c}}^\dagger_{0\sigma}  \hat{\tilde{c}}_{1\sigma} +\mathrm{h.c.} \rangle_\tau$. The required current and density-density correlation functions are experimentally accessible even in large systems~\cite{Impertro2024}. Measuring the correlators for~$N_{\rm init}$ initial states yields a $(N_{\rm init} \times 3)$-matrix~$M$, whose smallest singular value gives the best least square estimate of~$\{t,t',U\}$~\cite{Li2020}.

\begin{figure}[t!!]
\centering
\includegraphics[width=\linewidth]{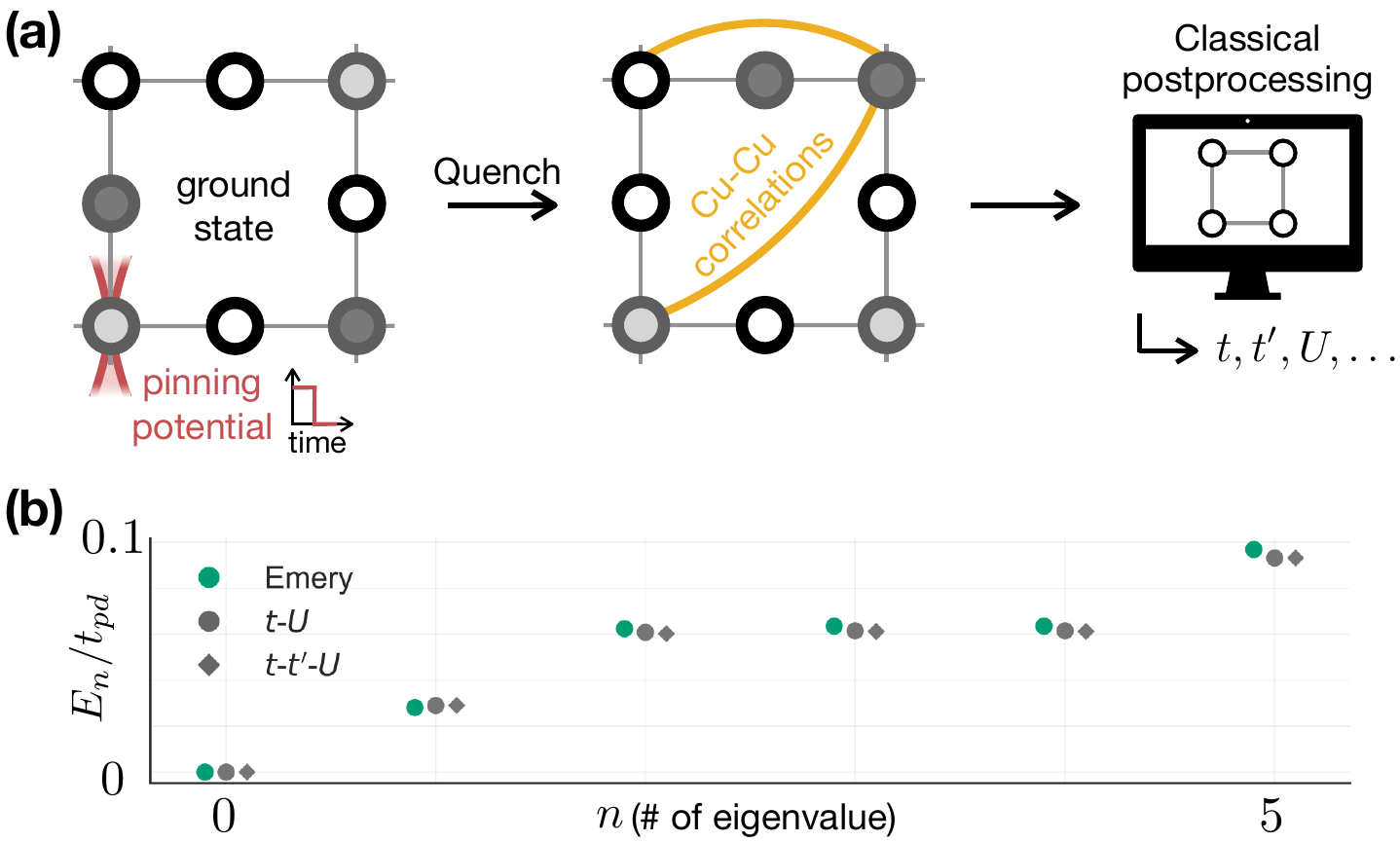}
\caption{\textbf{Learning effective single-band models.} \textbf{(a)} We prepare the system in the ground state of the three-band model, in the presence of small on-site potentials on Cu sites. Releasing those potentials, leads to a low-energy quench, where the system predominantly evolves within the low-energy manifold. Experimentally measuring the terms constituting an effective single-band Hubbard model, allows to learn such effective couplings (up to the absolute energy scale). \textbf{(b)} Using exact diagonalization, we compare the low-energy spectrum of the three-band model to learned, effective single-band models.   }
\label{figure-5}
\end{figure}

The choice of initial states -- fixing the energy density relative to the ground state of the three-band model -- must ensure that the subsequent dynamics remains confined to the low-energy sector. In contrast, an arbitrary initial state in the three-band model could result in substantial  occupation of the high-energy oxygen sites, that have no direct correspondence in the single-band model and hence obstruct a meaningful reconstruction~\cite{Jiang2023}. To this end, we initialize the system in the ground state of the three-band model subject to weak, random pinning potentials~$-\sum_{j_d} \varepsilon_{j_d} \hat{n}_{j_d}$ on the Cu sites with~$0 <\varepsilon_j \ll t_{ij}, \Delta_{pd}, U_\alpha$, see Fig.~\ref{figure-5}(a) left. The quench is induced by releasing the pinning potentials and evolving the system under the Emery Hamiltonian~\eqref{eq:Hamiltonian}. Inevitably, the quench gives rise to small oscillations between the low- and high-energy subspace (micromotion), which we mitigate by averaging over~$N_\tau$ time slices.

In our exact numerical simulations, we consider a single plaquette constituted by eight (four) sites in the three-band (single-band) model with $N_\uparrow=N_\downarrow=2$ fermions, see Fig.~\ref{figure-5}(a) left. We set $\Delta_{pd}/t_{pd} = 3.5$ and otherwise choose the same parameters as in Sec.~\ref{sec:DQMC}. First, we compute the ground state of the three-band model in the presence of pinning potentials~$\{\varepsilon_j\}$, sampled uniformly from~$[0,0.3t_{pd}]$. Then, we then time evolve the system up to time~$\tau \cdot t_{pd}=50$ in steps of~$\Delta \tau \cdot t_{pd}=4$. Afterwards, we compute the required correlators given by the terms in the \textit{single-band} model, i.e., between Cu-Cu sites, see Fig.~\ref{figure-5}(a) middle. We average over~$N_\tau=50$ time slices and repeat the process for~$N_{\rm init}=50$ initial states (distinguished by different~$\{\varepsilon_j\}$), to construct the matrix~$M$ and compute its smallest singular values, see Fig.~\ref{figure-5}(a) right.

To assess the quality of our protocol, we compare the low-energy spectrum of the three-band to the learned single-band model, see Fig.~\ref{figure-5}(b). We rescale and shift the spectra such that the ground-state energies~$E_0 \equiv 0$ in order to remove the ambiguity of absolute energy scale~\cite{Li2020}. We find qualitative agreement, indicating that the protocols finds an effective single-band model, which correctly captures the relevant low-energy features of the Emery model. The learned parameters are (i) $U/t\approx 16.7$ and (ii) $U/t\approx 16.7$ and $t'/t\approx -0.1$. In our small system and filling, we only find marginal effects of~$t'$, but we expect it to become more relevant in larger systems~\cite{Xu2024,Jiang2023}. We attribute deviations of our learned parameters from typical effective single-band models~\cite{Hirayama2018,Jiang2023} to finite-size effects; we highlight the agreement of the entire low-energy spectrum for the system size we consider.

An important direction for future work is to overcome several limitations of the present Hamiltonian learning protocol. There is a trade-off to choose sufficiently small quenches, in order to reduce micromotion between Cu and O sites, but injecting sufficient energy to induce dynamics on reasonably short timescales. Currently, we are mitigating the effect of micromotion by averaging over many initial realizations and evolution times, but more robust and efficient schemes have to be developed for experimental applications. In particular, the finite doping and smaller~$\Delta_{pd}$ regimes requires refined methods including the strong hybridization of $p$-$d$~orbitals~\cite{Eskes1993}.
We believe the learning of effective single-band Hubbard models from large-scale quantum quenches of the three-band model in experiments offers a new direction to find more accurate low-energy descriptions of high-$T_c$ superconductors.

\section{Outlook and Conclusions}
We have introduced an optical lattice architecture to realize Emery models with a wide tunability of parameters relevant to cuprate superconductors, including orbital-dependent Hubbard interactions. Our scheme combines bichromatic and phase-stabilized lattices, that both have been implemented previously~\cite{Tarruell2012,Xu2023,Chalopin2025}.
In the cuprate relevant regime, our architecture does not require potential engineering using a digital mirror device, reducing the risk of introducing disorder and finite couplings to other lattice sites.
To connect our theoretical proposal with experimental realizations, we have discussed signatures in quantum walks that characterize the band structure and studied the thermodynamics of the undoped system using large-scale numerical simulations. We predict that current quantum simulators can reach the charge-transfer insulating regime to probe the metal–insulator crossover; hence providing a platform for benchmarking numerical simulations of Emery models that suffer from double-counting corrections~\cite{Wang2012,Karolak2010,Vitali2019}. Finally, we proposed a Hamiltonian learning protocol to experimentally extract effective single-band Hubbard models -- central to the description of strongly correlated electrons -- from quantum quenches in the three-band model.

Our work narrows the gap between cold-atom quantum simulators and strongly-correlated materials.
Experimental implementations at low temperatures will allow the microscopic exploration of the interplay between orbitals within the phase diagram of cuprate and nickelate superconductors as well as other transition-metal oxides.
Large-scale numerical simulations will play an important role to guide and benchmark experiments. Our architecture naturally enables the realization of additional terms that might play a significant role in high-$T_c$ materials, such as extended Hubbard interactions~$U_{pd}$~\cite{Hansmann2014}, for example by employing atoms with magnetic dipole moments or dipolar molecules~\cite{Carroll2025}.
\\ \\
During completion of this manuscript, we became aware of a closely related work proposing a single-color, interfering optical lattice to implement Emery models~\cite{Lange2026}.

\section*{Acknowledgements}
We thank Annette Carroll, Matja\v{z} Kebri\v{c} and Lode Pollet for valuable feedback on the manuscript, and Lev Kendrick, Youqi Gang, Alex Deters, Aaron Young, Andrew Millis, Shiwei Zhang, Benjamin Cohen-Stead for fruitful discussions. AMR and LH acknowledge support by the Simons Collaboration on Ultra-Quantum Matter, which is a grant from the Simons Foundation (651440). JB acknowledges support from the NSF Graduate Research Fellowship Program. This project is supported by the NSF JILA-PFC PHY-2317149, the ARO Grant W911NF2410212, and the Gordon and Betty Moore Foundation under grant DOI 10.37807/GBMF12948. 

%\pagebreak
%\clearpage
%\newpage

\appendix
\setcounter{section}{0}

%\onecolumngrid

%\begin{center} 
%\textbf{\large End matter}
%\end{center}

\section{Sign pattern of tunneling amplitudes} \label{app:gauge-trafo}
The phase structure of $d_{x^2-y^2}$ and $p_{x,y}$ orbitals is reflected in the sign pattern of the tunneling amplitudes in transition-metal oxides. Since there is a gauge freedom in choosing the phases of local orbitals, a natural choice for these materials is a gauge in which the orbitals are translationally invariant, which we refer to as ``material gauge''. In contrast, the tunneling amplitudes between optical lattice sites reflect the rotational $C_4$ symmetry of the lattice potential. Therefore, we apply a gauge transformation to demonstrate the equivalence between our tight-binding description of the optical lattice implementation, Eq.~\eqref{eq:Hamiltonian}, and the Emery model in the material gauge.

\begin{figure}[t!!]
\centering
\includegraphics[width=\linewidth]{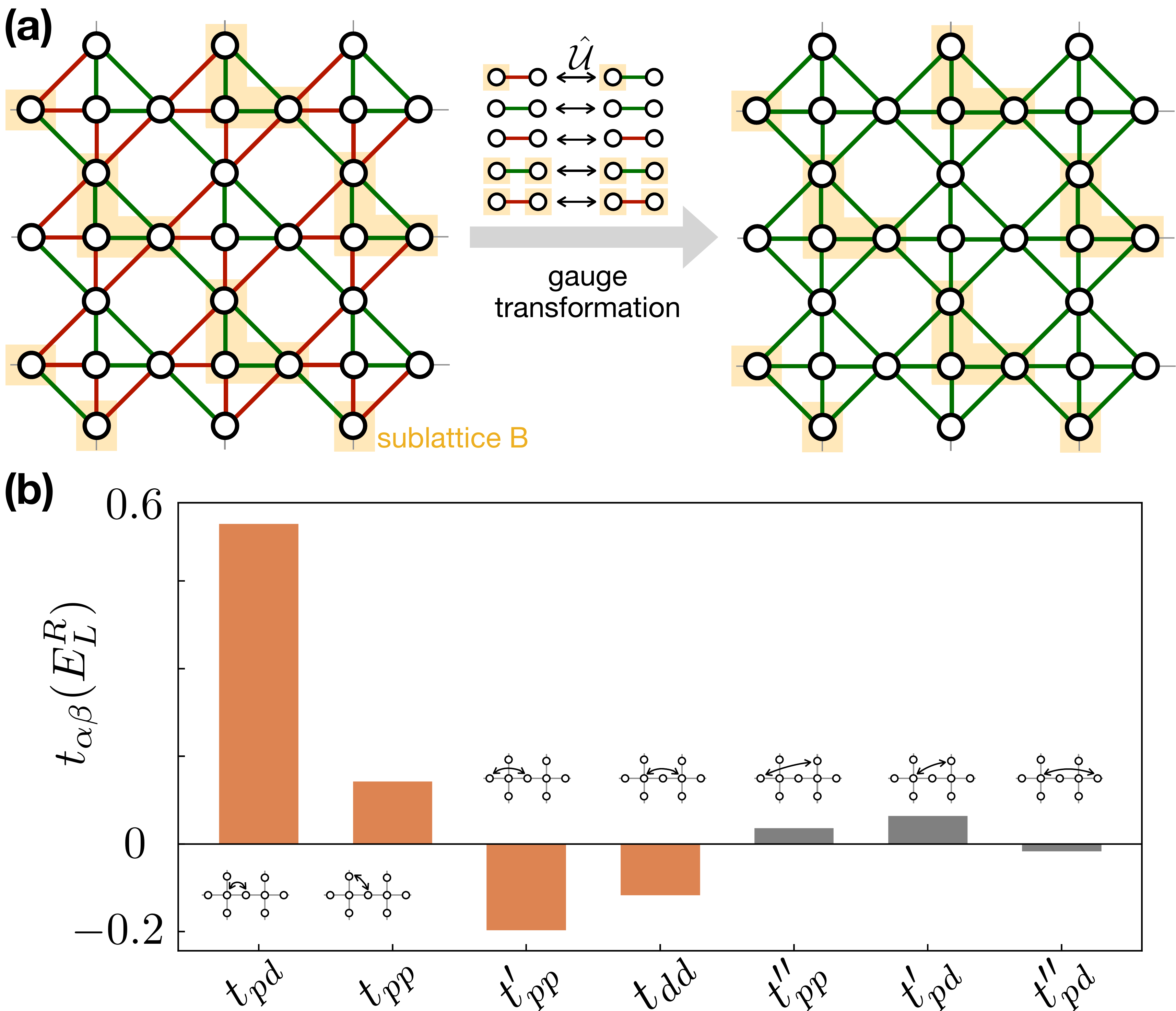}
\caption{\textbf{Tunneling amplitudes.} \textbf{(a)} Left: The sign structure of tunnelings~$t_{pd}$ and~$t_{pp}$ has negative (red) and positive (green) amplitudes, arising from the symmetry of the~$d_{x^2-y^2}$ and $p_{x,y}$ orbital. Right: By applying a gauge transformation~$\hat{\mathcal{U}}$ that acts on fermions in sublattice B as $\hat{\mathcal{U}}^\dagger \hat{c}_{j_B}\hat{\mathcal{U}} = -\hat{c}_{j_B}$, the sign of tunnelings becomes uniform. Similarly, the signs of the all other tunnelings considered in our tight-binding model are modified accordingly. \textbf{(b)} Sign structure of tunneling amplitudes obtained \textit{ab initio} from optical lattice Wannier functions, consistent with the band structure in cuprates~\cite{Kent2008} after the gauge transformation~$\hat{\mathcal{U}}$. The tunneling amplitudes in orange are included in the DQMC calculations. }
\label{figure-gauge-trafo}
\end{figure}

First, we note that the Lieb lattice is bipartite, i.e., the unit cells form a checkerboard structure with sublattices~$A$ and~$B$, each containing three sites, as indicated by the yellow boxes in Fig.~\ref{figure-gauge-trafo}(a). We then define a gauge transformation~$\hat{\mathcal{U}} = \exp(i\pi\sum_{j \in j_B}\hat{n}_{j})$, such that
\begin{align}
    \hat{\mathcal{U}}^\dagger \hat{c}_j^{(\dagger)}\hat{\mathcal{U}} = \begin{cases} \hat{c}_j^{(\dagger)} & \mathrm{for~} j\in j_A \\ -\hat{c}_j^{(\dagger)} & \mathrm{for~} j\in j_B.\end{cases}
\end{align}
This transformation acts non-trivially on tunneling terms between adjacent unit cells, $-t_{j_Aj_B} \hat{c}_{j_A}^{\dagger}\hat{c}_{j_B} \rightarrow +t_{j_Aj_B}\hat{c}_{j_A}^{\dagger}\hat{c}_{j_B}$. As shown in Fig.~\ref{figure-gauge-trafo}(a), applying this transformation to the Emery model in the material gauge leads to $C_4$ symmetric tunneling amplitudes, as required for its implementation in optical lattices.

Similarly, the gauge transformation modifies all other tunneling terms we consider in our tight-binding description. We note that including tunneling terms between next-nearest-neighbor unit cells would introduce sign structures that cannot be removed by a gauge transformation. However, as discussed in Fig.~\ref{figure-3}(a), tunnelings up to nearest-neighbor unit cells are sufficient to obtain quantitative agreement with the optical lattice band structure. Hence, the resulting sign structure of the tunneling amplitudes is consistent with that of cuprates, see Fig.~\ref{figure-gauge-trafo}(b). 

Finally, we emphasize that observables must be interpreted with respect to the chosen gauge when comparing to the material gauge. This is important for observables that transform non-trivially under~$\hat{\mathcal{U}}$, such as current operators or single-particle Green's functions.

%%%Optical lattice
\section{Optical lattice calculations} \label{app:Wannier}

The tight-binding parameters of Hamiltonian~\eqref{eq:Hamiltonian} are obtained \textit{ab initio} from the optical lattice potential. To obtain the Wannier functions of our three-band model, we perform a steepest-descent algorithm to find the maximally-localized Wannier functions (MLWF) \cite{Vanderbilt2012} for each set of parameters~$V_L, V_S, \phi_S$.

\begin{figure}[t!]
    \centering
\includegraphics[width=\linewidth]{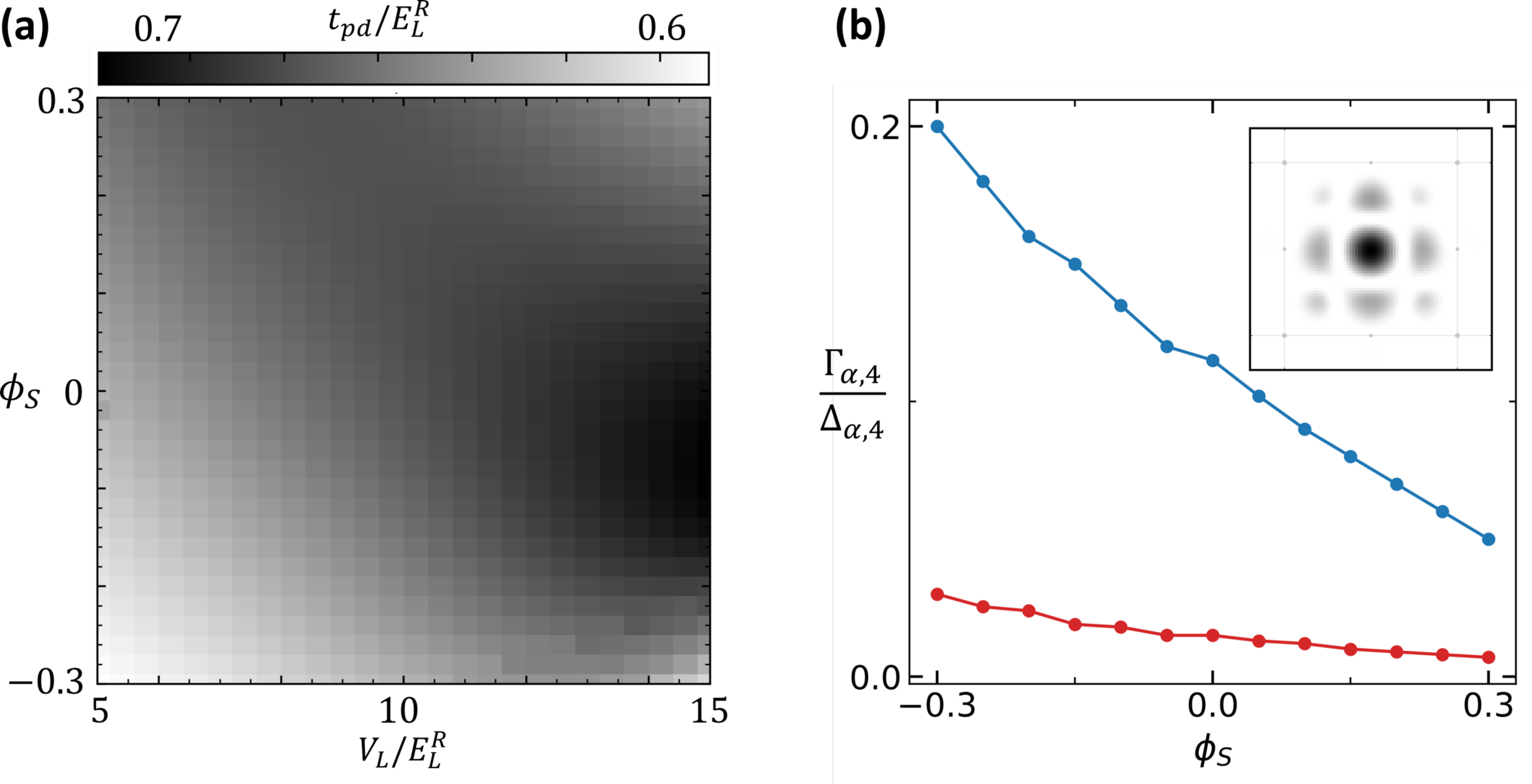}
    \caption{\textbf{Tight-binding calculations.} \textbf{(a)} The nearest-neighbor tunneling $t_{pd}$ in recoil units up to $V_{L} = 15E^R_L$. As trap depth increases at constant $\phi_S$, the offset $\Delta_{pd}$  grows correspondingly, causing the $p$-orbital Wannier function to spread further. This expanded spatial extent enables enhanced tunneling to deeper trap depths. However at larger $V_{L} \geq 15V^R_L$ (not shown)  all tunnelings decay exponentially. \textbf{(b)} The perturbative admixing, induced by atomic collisions, of the fourth band with the Emery bands (red for $\alpha=d$ and blue for $\alpha =p$) for $V_L = 10E^R_L$ as we vary the interference angle. The inset shows the Wannier function of the fourth band which is located at the ``fake site".}
    \label{fig:tight-binding}
\end{figure}

\textit{Algorithm.---}First, the Bloch eigenstates $\psi_{\mathbf{k}}^{(m)}(\mathbf{r})$ ($m$ the band index) are obtained by numerically solving the single-particle Schr{\"o}dinger equation for the optical lattice potential. 
The Wannier functions are found from the eigenstates~$\psi_{\mathbf{k}}^{(m)}(\mathbf{r})$ of the optical lattice potential through
\begin{align}
    w^{(n)}_{j}(\mathbf{R}) = \frac{\Upsilon}{(2\pi)^D}\int _{BZ}d\mathbf{k} \sum_{m=1}^{J}M^{\mathbf{k}}_{nm}\psi^{(m)}_{\mathbf{k}}(\mathbf{r})e^{-i\mathbf{k}\cdot \mathbf{R}}
\end{align}
where $n,m$ label the band index, $\mathbf{k}$ is the momentum, and $\mathbf{r}$ the spatial coordinates. $\Upsilon$ is the volume of the primitive cell of the $D-$dimensional direct lattice (in our case $2\pi$).    
Finding the set of Wannier functions amounts to choosing  $M^{\mathbf{k}}_{nm}$, a $\mathbf{k}$-dependent $3\times 3$ matrix which mixes the bands, such that they are maximally localized on each site. To do this, we apply the methods of Refs.~\cite{Walters2013,Marzari1997}, which we briefly summarize here. The unitary $M$ is iteratively optimized until the spread $\Omega = \sum_{n} \langle \mathbf{r}^2 \rangle_n - \langle \mathbf{r}\rangle^2_n $ of the orbitals is minimized. The resulting Wannier functions are termed \textit{ordinary} or \textit{generalized} depending on whether $M$ was chosen to be diagonal or to contain off-diagonal entries, respectively.  

We work in a truncated Fourier space such that reciprocal lattice vectors with magnitudes larger than $G_{\rm max} = 100/\lambda_{L}$ are neglected. This limits spatial resolution to $2\pi/G_{\rm max} = \pi \lambda_{L}/50$, which is a length scale longer than that over which the optical lattice profile changes significantly. 
Integrals over a primitive cell of $\mathbf{k}$-space are replaced by a sum over a discrete mesh of wave-vectors. Specifically, the mesh is formed by the wave-vectors $\mathbf{k} = \mathbf{G}/N$ within some primitive cell of the reciprocal lattice, where $\mathbf{G}$ is a reciprocal lattice vector. This is equivalent to working with a periodic system with only $N$ primitive cells in each dimension, and therefore strictly the relation $N \gg  1$ should be satisfied in order for boundary effects to be negligible. In our case, we use $N = 20$ for all calculations.

First, we find the band structure and from that, the Bloch states are calculated.   Relative to the Bloch states, a gauge corresponding to the set of ordinary MLWG is found. This amounts to first optimizing over the diagonal entries of the $M^{\mathbf{k}}_{nm} = e^{i\phi_{n}(\mathbf{k})}\delta_{nm}$  matrix. 
We then find a gauge (relative to the same initial Bloch states) corresponding to a set of generalized Wannier states with a reduced inter-band contribution to the spread. Immediately after this, the gauge is transformed such that the intra-band contribution to the spread is minimized (the inter-band contribution is unaltered). This corresponds to optimizing over the off-diagonal entries of $M^{\mathbf{k}}$. 
For each stage, the optimization is performed via steepest-descent minimization using an infinitesimal unitary step $M \rightarrow M + \epsilon_{\rm step}dM$. 
The spread functional $\Omega = \Omega_{I} + \Omega_{D} + \Omega_{OD}$ can be decomposed into three terms; one invariant under the gauge $\Omega_{I}$, one only dependent on the diagonal entries of the unitary $\Omega_{D}$ and finally one only dependent on the off-diagonal entries of the unitary $\Omega_{OD}$. Hence we can independently optimize the ordinary Wannier functions before moving onto the generalized Wannier functions via band mixing.

Around $N_{\rm iter} = 4000$ iteration steps are used for the steepest-descent minimization of the intra-band spread for a single band, the disentangling procedure to reduce the inter-band spread and the steepest-descent minimization of the total spread for a single band. 
For each reported value, these steps result in GMLWFs with spreading $\Omega \lesssim 10^{-2}\lambda_L$.
While $\Omega$ provides a convenient functional to minimize over starting from non-interacting states, one can obtain tight-binding parameters starting from two-particle Wannier functions~\cite{Busch2017}. This approach leads to a smaller onsite interaction at shallower trap depths relative to the single-particle approach. As our $p$-site is at a lower effective trap depth than the $d-$sites, we expect that the interaction ratio $U_{d}/U_{p}$ would only increase if a two-particle approach was followed, and hence more realistic for cuprate parameters.

\textit{Tight-binding parameters.---}From these set of Wannier functions, we can find the tight-binding parameters via 
\begin{align}
    t_{\mathbf{R}\mathbf{R'}}^{nm} &= - \int d\mathbf{r} w^{* (n)}_{\mathbf{R}}(\mathbf{r})\hat{h}w^{(m)}_{\mathbf{R'}}(\mathbf{r})  \\
    U^{nm}_{\mathbf{R}\mathbf{R}'} &= g \int d\mathbf{r} |w^{(n)}_{\mathbf{R}}(\mathbf{r})|^2 |w^{(m)}_{\mathbf{R}'}(\mathbf{r})|^2
\end{align}
where $\hat{h} = \hat{\mathbf{p}}^2/2m + V(\hat{\mathbf{r}})$ is the single-particle Hamiltonian and we consider a contact pseudopotential such that $g = \frac{2a_s}{\pi\lambda_S} E^{R}_{L}$, where $a_s$ is the $\mathrm{SU}(2)$-invariant scattering length of the fermionic atoms in their pseudospin states, tunable via atomic scattering resonances in~$^{6}\mathrm{Li}$. The indices $\mathbf{R},\mathbf{R'}$ sets the location of the unit cells.

In Fig.~\ref{fig:tight-binding}(a), we plot the corresponding absolute tunneling scale of~$t_{pd}$ for the relevant parameter regime discussed in the main text in Fig.~\ref{figure-2}(b).

\textit{Heating rate.---}If we prepare a sample in the three-lowest bands of our lattice, there is a scattering rate, induced by collisions, to excite atoms into a higher band~$m$, which eventually determines the feasibility of our scheme. In the perturbative regime, if this collision rate $\Gamma_{n,m}$ (calculated via $U^{nm}_{00}$ as shown above) which couples a low-energy band~$n=1,\ldots,3$ to the higher bands~$m\geq4$ is much smaller than the energy gap, $\Gamma_{n,m} < \Delta_{n,m}$, then we can restrict the dynamics to a three-band Hubbard model. We plot this metric for the $d$ and $p$ Wannier functions to the $m=4$ band in Fig.~\ref{fig:tight-binding}(b).

For $V_S/V_L = 1.25$ and below $\phi_S = -0.2$, the band of the fake site approaches the Emery bands and the leakage increases, see Fig.~\ref{fig:tight-binding}(b), for which we propose projecting repulsive potentials with a digital mirror device to mitigate occupying the fourth band~\cite{Lebrat2024_Lieb}. We highlight that this is beyond the cuprate relevant regime. Further, as the cuprate regime corresponds to $\Delta_{pd}/t_{pd} \geq 3$, where the occupation in the $p$-band is quite small, it is reasonable to expect the true leakage rate would be further reduced.
In principle there can be offsite Hubbard interactions or density-assisted tunnelings between the Emery bands. We have found these terms to be negligible. 

\begin{figure}
    \centering
    \includegraphics[width=\linewidth]{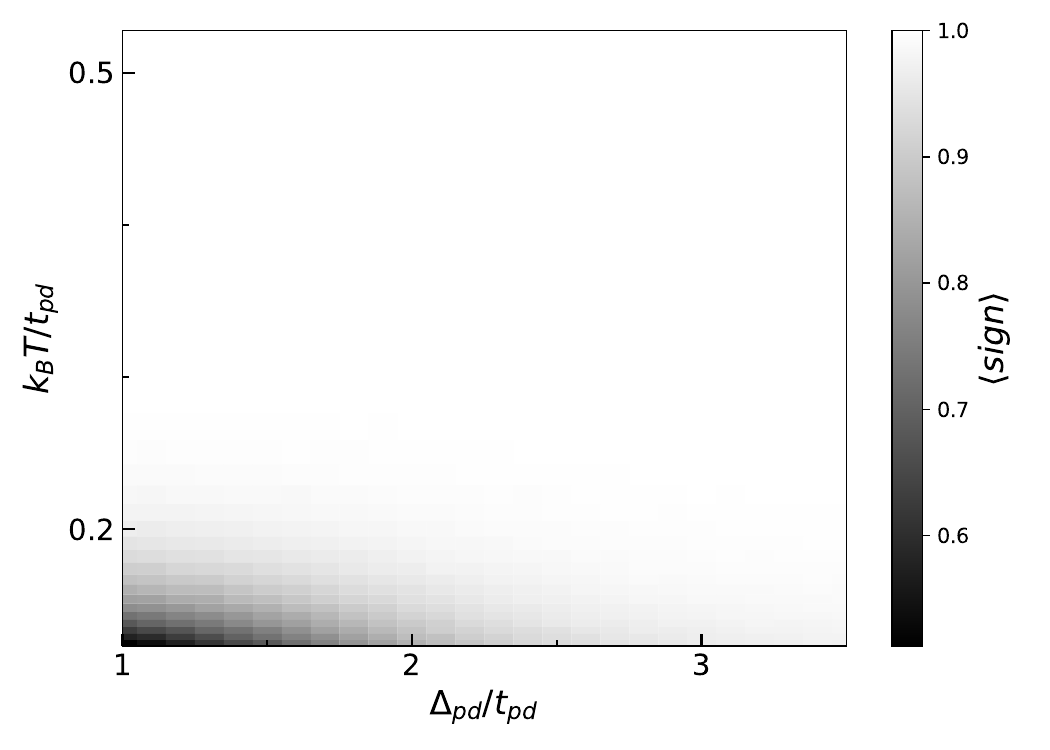}
    \caption{\textbf{DQMC sign problem.} The sign error for $L=8$, as a function of temperature and offset. We encounter a sign problem at low temperatures for $\Delta_{pd}/t_{pd} < 1$ due to our inclusion of finite $t_{pp}$ and $U_{p}$ terms.}
    \label{fig:sign}
\end{figure}
%%%DQMC
\section{Determinant Quantum Monte Carlo}
The Determinant Quantum Monte Carlo (DQMC) results were obtained using the \texttt{SmoQyDQMC} package v2.0~\cite{SmoQyDQMC.jl,SmoQyDQMC.jl_Codebase}. The tight-binding parameters used were $U_{d}/t_{pd} = 7$, $U_{p}/t_{pd} = 3.5$, $t_{pp}/t_{pd} = 0.2$, $t_{pp}'/t_{pd} = -0.1$, $t_{dd}/t_{pd} = -0.1$.
The required doping of $n= 1/3$ was achieved by dynamically adjusting the chemical potential during the simulation through the method developed in Ref.~\cite{Miles2022}, incorporated into the package. For instance, as the chemical potential is intensive and weakly depends on system size, we run the DQMC for a smaller system $L=4 \times 4$ to find the final chemical potentials for each parameter and use those final values as initial estimates for the larger system sizes which speeds up convergence.  
To thermalize, we use $N_{\rm therm} = 5000$ updates. We take $N_{m} = 5000$ measurements distributed across $N_{\rm bins} = 40$ with $125$~measurements each. The number of updates between bins was $N_{\rm updates} = 5$.

For each point $(k_BT/t_{pd},\Delta_{pd}/t_{pd})$, we use $N_{w} = 8$ independent walkers, running their own independent DQMC simulation each with independent random initial seeds. Each reported data point is obtained by averaging across all walkers. This is useful for the spin correlations: while the Emery model is ${\rm SU}(2)$-symmetric and hence measurements of $ \hat{\mathbf{S}}\cdot \hat{\mathbf{S}} $ and $\hat{S}^z \hat{S}^z $ should provide the same information, it is possible for the thermalized state to get stuck in a given magnetization direction. 

We compute the thermodynamic compressibility and spin-spin correlation functions
where the compressibility is averaged over the full number of sites whereas the spin-spin correlations are averaged over the number of unit cells.

\begin{figure}[t!!]
    \centering
    \includegraphics[width=\linewidth]{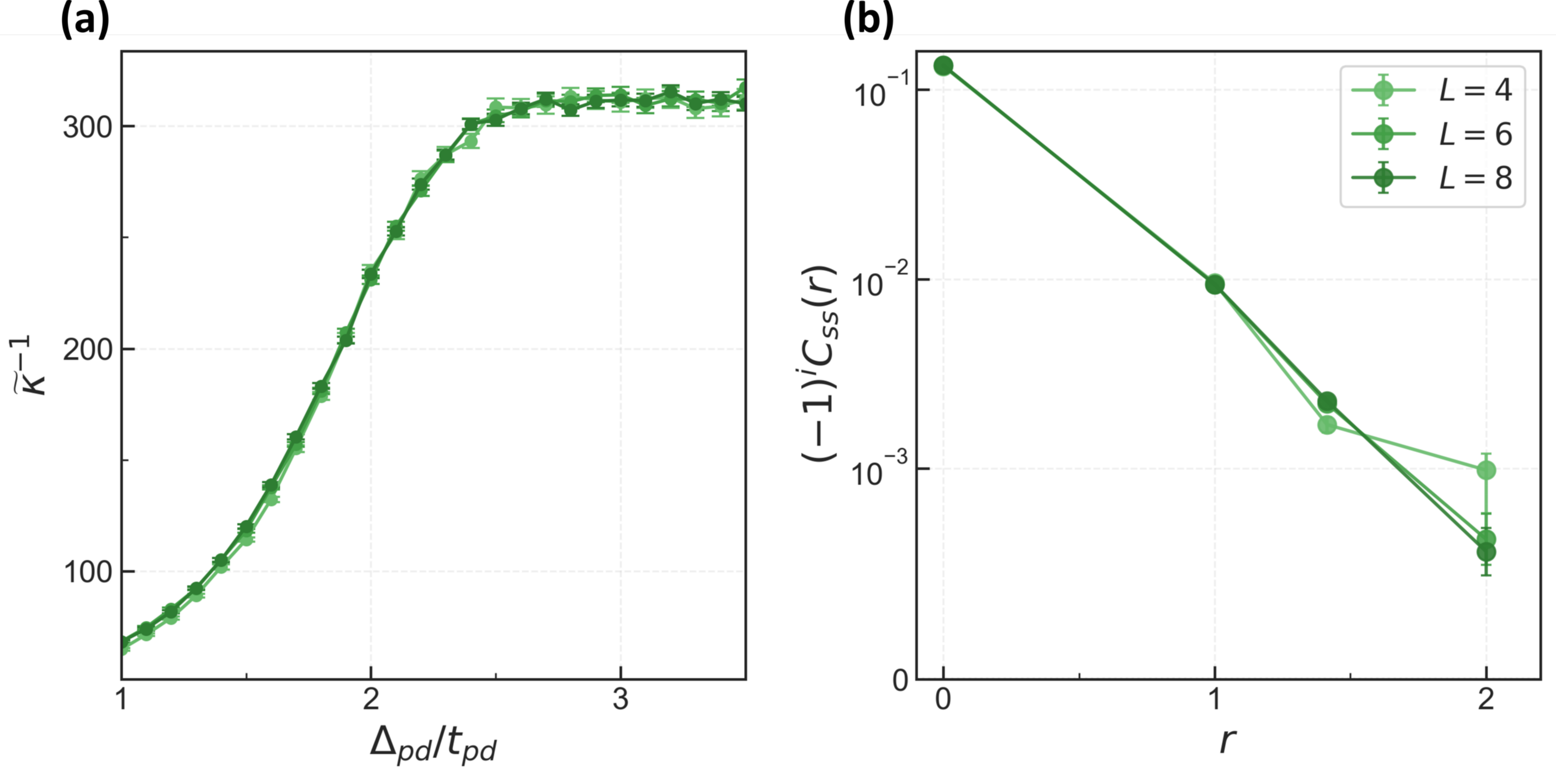}
    \caption{\textbf{Convergence of DQMC data.} \textbf{(a)} The dimensionless incompressibility~$\tilde{\kappa}^{-1}$ at the lowest temperature we consider $k_B T/t_{pd} = 1/8$ as a function of $\Delta_{pd}/t_{pd}$ for different system sizes. \textbf{(b)} The staggered spin-spin correlations for $\Delta_{pd}/t_{pd}=1$ as a function of distance~$r$. The cutoff of the semilog scale is set to $|10^{-3}|$, i.e., the scale is linear for~$\leq |10^{-3}|$. }
    \label{fig:dqmc}
\end{figure}
In DQMC, integrating out fermions leads to configuration weights given by fermion determinants, which are not guaranteed to be positive. Since Monte Carlo sampling requires non-negative probabilities, one samples using the absolute value of the determinant and includes its sign in observables. This results in an exponential growth of statistical uncertainties in the so-called $\textit{sign problem}$.
For low enough temperatures, we get a sign error for smaller offsets $\Delta_{pd}/t_{pd} \sim 1$, see Fig.~\ref{fig:sign}. Previous work has found that the introduction of $t_{pp}$ and $U_{p}$ results in a sign problem even at half-filling~\cite{Devereaux2016}. As $\Delta_{pd}/t_{pd}$ increases the $p$-occupation substantially reduces, see Fig.~\ref{figure-4}, these terms become less relevant resulting in the sign returning to unity. 

We run the calculations for $L \times L$ unit cells with periodic boundary conditions where $L=4,6,8$ (corresponding to $\mathcal{N} = 48, 108,192$ sites) and find system-size convergence at $L=6$, see Fig. \ref{fig:dqmc}.
We choose $L$ even such that any antiferromagnetic order is commensurate with the periodic boundary conditions as using odd $L$ might lead to frustration in small systems. While standard errors for all values of the compressibility remain low ($<1\%$), the spin-spin correlations display insufficient large signal-to-noise past $r>2$. 

\begin{figure}[b!!]
    \centering
    \includegraphics[width=\linewidth]{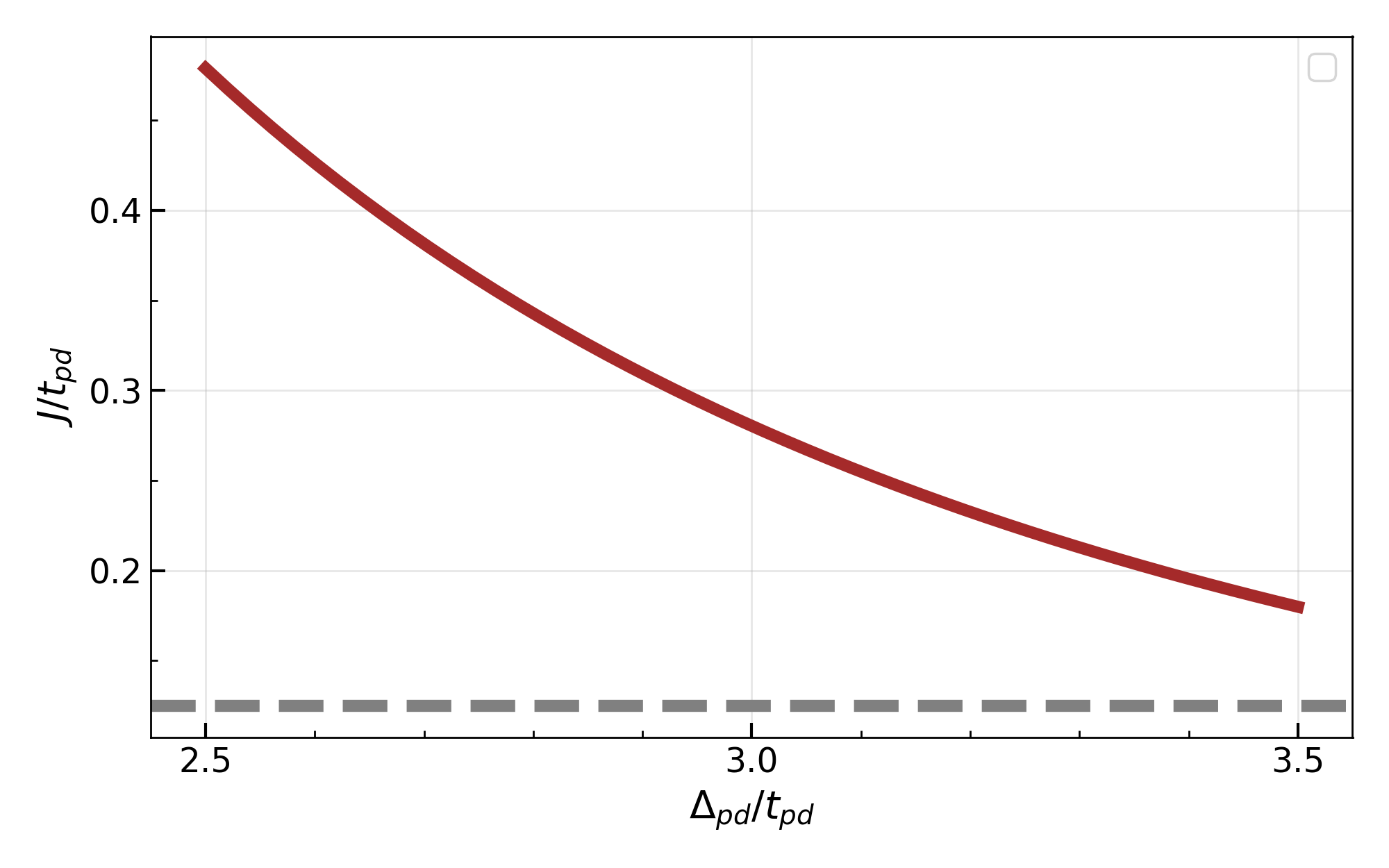}
    \caption{\textbf{Emery model superexchange scale.}  The dashed line indicates the lowest temperature $k_B T/t_{pd} = 1/8$ we consider in DQMC.}
    \label{fig:superexchange}
\end{figure}
%%%Superexchange
\section{Superexchange in the three-band model}
\label{app:superexchange}
For $U_d, \Delta_{pd}, U_{p} \gg t_{\alpha \beta}$, the superexchange between $d$-sites in the undoped system can be represented by an effective Heisenberg Hamiltonian $\hat{H}_J = J\sum_{\langle i_d, j_d \rangle}\left( \mathbf{S}_{i_d}\cdot \mathbf{S}_{j_d} - \frac{1}{4}\hat{n}_{i_d}\hat{n}_{j_d} \right)$ with the scale set as~\cite{Eskes1993}: 

\begin{equation} \label{eq1}
\begin{split}
J = &\frac{4t_{pd}^4}{\Delta_{pd}^2}\Bigg{(} \frac{1}{U_d} + \frac{2}{2\Delta_{pd} + U_p} + \frac{8t_{pp}}{U_d\Delta_{pd}} \\ &+ \frac{16t_{pp}}{\Delta_{pd}(2\Delta_{pd} + U_p)} + \frac{4t_{pp}}{\Delta_{pd}^2} \Bigg{)}
\end{split}
\end{equation}
which we plot in Fig.~\ref{fig:superexchange} with the parameters we used for DQMC indicating our lowest considered temperature is below the superexchange scale. We neglect contributions from $t_{dd} \ll t_{pd}$ (which arise at second and third order), however these should just raise the scale further.

\providecommand{\noopsort}[1]{}\providecommand{\singleletter}[1]{#1}%

\end{document}